\documentclass[twoside]{dis09}
\usepackage[latin1]{inputenc}
\usepackage[dvips]{graphicx,epsfig,color}
\usepackage{wrapfig,rotating}
\usepackage{amssymb,amsmath,array}

\begin{document}
\title{Supersymmetry}

%***********************************************************************
% AUTHORS INFORMATION AREA
%***********************************************************************
\author{Dirk Zerwas
%
% Optional short acknowledgment: remove next line if non-needed
%\thanks{This is an optional funding source acknowledgment.}
%
% DO NOT MODIFY THE FOLLOWING '\vspace' ARGUMENT
\vspace{.3cm}\\
%
% Addresses and institutions (remove "1- " in case of a single institution)
Laboratoire de l'Acc\'el\'erateur Lin\'eaire, IN2P3/CNRS,\\
Batiment 200, 91898 Orsay Cedex, France 
%
% Remove the next three lines in case of a single institution
}
%***********************************************************************
% END OF AUTHORS INFORMATION AREA
%***********************************************************************

\maketitle

\begin{abstract}
The determination of supersymmetric parameters at the LHC
in favorable as well as difficult scenarios is presented. If discovered
and measured at the LHC and the ILC, supersymmetry
may provide a link between collider physics and cosmology. 
\end{abstract}

\section{Introduction}

In this report based on the talk~\cite{url}, elements of the status of supersymmetry are discussed. 
In a supersymmetric theory~\cite{Martin:1997ns}
each fermionic degree of freedom has a bosonic counter
part and vice versa. Supersymmetry has no problems with radiative corrections
due to the cancellation of quadratic divergences. A light Higgs boson with a mass
of less than about 150~GeV is predicted. Supersymmetry paves a path to including gravity in the
theory of elementary particles. 

The particle spectrum of a supersymmetric theory consists of at least three
neutral Higgs bosons, two or more charged Higgs bosons and supersymmetric particles
such as the sleptons, squarks, the neutralinos, the charginos and the gluino.

\begin{wrapfigure}{r}{0.5\columnwidth}
\vspace{-0.75cm}
\includegraphics[width=0.45\columnwidth]{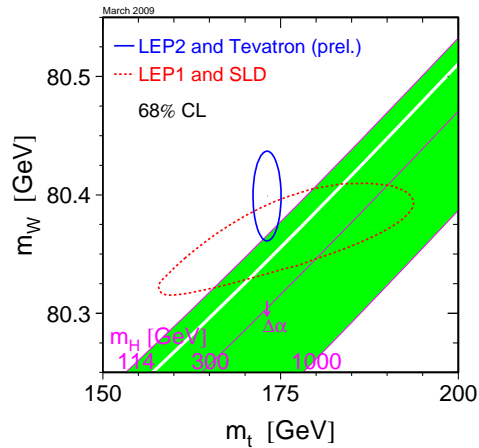}
\vspace{-0.5cm}
\caption{The error ellipse of the top quark and W~boson mass measurements at LEP and TeVatron
is shown and the lines for different Higgs boson masses is indicated.
(update of~\cite{Alcaraz:2007ri}).}\label{Fig:MWMT}
\end{wrapfigure}

The origin of supersymmetry breaking is not known today. Several models
parametrize this ignorance, among them: the minimal supersymmetric extension of the 
standard model (MSSM), minimal supergravity (mSUGRA), decoupled scalars supersymmetry (DSS).
Other models such as the NMSSM, the MRSSM and the N=1/N=2 hybrid model
add further fields (particles) to the minimal particle content of the MSSM.

R--parity is assumed to be conserved in the following. Thus 
supersymmetric particles are produced in pairs and cascade decay to the lightest supersymmetric 
particle (LSP). The LSP is stable, neutral and weakly interacting, for the lightest neutralino 
will be used.
The general experimental signature for the detection of supersymmetry at colliders is missing
transverse energy.

Supersymmetry stabilizes
the gap between the electroweak and Planck scales, unifies the couplings of the three forces,
can induce electroweak symmetry breaking, provides a link to physics at the GUT scale and 
a candidate for dark matter. Experimentally the electroweak data favor a light Higgs boson,
compatible with supersymmetry as shown in Figure~\ref{Fig:MWMT} for the W~boson mass as function
of the top quark mass. 

\section{Vintage and Difficult Supersymmetry}

\begin{wrapfigure}{r}{0.5\columnwidth}
\vspace{-0.75cm}
\centerline{\includegraphics[width=0.45\columnwidth]{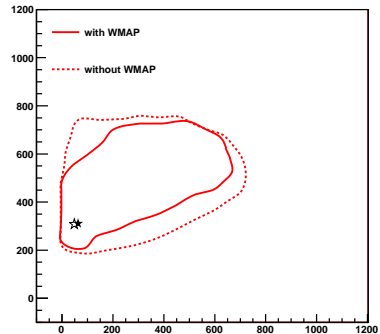}}
\vspace{-0.5cm}
\caption{The best fitting point (star) determined from the electroweak fit (plus
additional observables) is shown in the (m$_0$,m$_{1/2}$) plane with 
the 1$\sigma$ contour~\cite{Buchmueller:2008qe}.}\label{Fig:SPS1aIndirect}
\end{wrapfigure}

A typical parameter set for vintage mSUGRA studies is SPS1a~\cite{Allanach:2002nj}.
SPS1a has moderately heavy gluinos and squarks, heavy and light gauginos and the lightest Higgs boson
at the LEP limit. Important for the LHC, the long cascade 
$\tilde{q}_L\rightarrow\chi_2^0 q\rightarrow\tilde{\ell}_R\ell q\rightarrow\ell\ell q \chi^0_1$
is detectable~\cite{Weiglein:2004hn}. While this choice of parameters was initially thought
to be optimistic, it has been shown~\cite{Buchmueller:2008qe} that today's electroweak precision data
together with the relic density measurement and rare b branching ratios
result in a best--fit mSUGRA parameter point quite close to SPS1a as shown in Figure~\ref{Fig:SPS1aIndirect}.
The parameter set SPS1a has been analyzed in detail for measurements at the LHC and ILC. At the LHC 
the squark sector, gluinos, sleptons and part of the neutralino sector can be discovered. Edges and thresholds 
are observed in invariant mass distributions. These are functions of the intervening masses and do
not depend on the underlying theory. As soon as the production threshold is passed at the ILC, 
the particle can be measured.

Transforming the measurements into a determination of the underlying parameters 
is a formidable task for which sophisticated tools are necessary.
An overview of the tools for supersymmetric predictions such as masses, production 
cross sections and branching ratios can be found in~\cite{Allanach:2008zn}.
Fittino~\cite{Bechtle:2005vt} and SFitter~\cite{Lafaye:2004cn,Lafaye:2007vs} provide
different techniques to sample a multi-dimensional parameter space with correlated
measurements. Weighted Markov chains can be used for efficient sampling 
of a high dimensional parameter space as it is essentially linear in the number 
of parameters. After having produced a full--dimensional exclusive likelihood 
map, two options of projections can be explored: marginalization (Bayesian), which introduces
a measure, or the profile likelihood (frequentist) approach of choosing the parameter 
with the maximum log-likelihood value.

A first example is the determination of the mSUGRA parameters as an illustration
of a model with few parameters which are (mostly) defined at the GUT scale. 
The correct parameter set
is found from the measurements, but the Markov chains also locate secondary minima in this
tightly constrained scenario. One of the secondary minima results from the interplay of the
top quark mass and the trilinear coupling A$_0$. It is important that the standard model 
parameters are determined coherently including their experimental errors.
The secondary minima can be eliminated by comparing the log-likelihood values which are larger
for the correct solution.

After having determined the central values of the parameters, the errors on the parameters have to be determined
either in a single fit or by using toy experiments. Theory errors are considered to be flat
in contrast to the experimental errors which are treated as Gaussian and can have strong correlations
in the systematic part. With this rigorous treatment of the errors, the expected errors at the LHC are of the
order of percent or better and are improved by a factor of three or four when the expected results from the ILC 
are included. The inclusion of theory errors in the analysis has an impact on the precision not only 
on the ILC errors, but also on the LHC errors (precision is decreased approximately by a factor three).

\begin{wrapfigure}{r}{0.5\columnwidth}
\vspace{-0.75cm}
\centerline{\includegraphics[width=0.45\columnwidth]{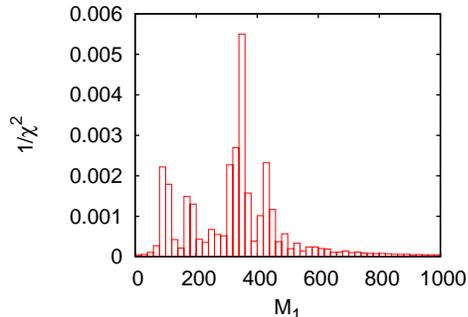}}
\vspace{-0.5cm}
\caption{The distribution of the gaugino mass parameter $M_1$ is shown. 
Several secondary minima are observed~\cite{Lafaye:2007vs}.}\label{Fig:SFitterM1}
\end{wrapfigure}

The MSSM is defined at the electroweak scale with up to about 120 parameters. Including
phenomenological constraints (absence of FCNC etc), 
the parameters can be reduced to 19~parameters. The larger parameter space
necessitates a thorough search for secondary minima. Only three gaugino masses can be 
observed at the LHC in SPS1a, in the MSSM the gaugino/Higgsino sector is therefore
under-constrained resulting in a eightfold ambiguity (see Figure~\ref{Fig:SFitterM1}) 
in this sector where the log-likelihood
values are essentially identical. Adding the ILC measurements to the analysis, all parameters 
can be determined and the ambiguities can be resolved. The MSSM also has the particular advantage that
once the parameters have been determined, they can be extrapolated to the high scale, thus 
measuring grand unification instead of imposing it as shown by~\cite{Blair:2002pg}.

Once the supersymmetric parameters have been determined either in mSUGRA or the MSSM, the
full spectrum and the couplings can be deduced. Calculating 
the relic density from this information, a collider prediction for $\Omega$h$^2$ is obtained~\cite{Baltz:2006fm}.
In the parameter set SPS1a from LHC data alone, ignoring theory errors, a precision of about 2\% can be 
obtained, adding ILC measurements the precision would improve by an order of magnitude. The precision,
albeit depending strongly on the exact phenomenology of supersymmetry, is comparable to that 
of WMAP and the expectation of Planck. Confronting these two measurements will provide for interesting
studies in the future.

\section{Models and the Higgs sector}

\begin{wrapfigure}{r}{0.5\columnwidth}
%\vspace{-0.75cm}
\centerline{\includegraphics[width=0.35\columnwidth,angle=-90]{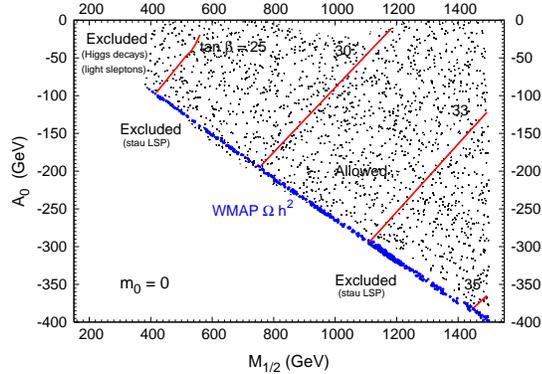}}
\vspace{-0.3cm}
\caption{After applying all constraints, including the relic density, only the approximately diagonal line
in the (m$_{1/2}$,A$_0$) plane is allowed for small m$_0$~\cite{Djouadi:2008yj}.}\label{Fig:cNMSSM}
\end{wrapfigure}

In the NMSSM a singlet is added to the MSSM in order to generate the $\mu$~term
dynamically. The Higgs 
sector is extended to five neutral Higgs boson and an additional neutralino is present. 
The price to be paid are additional parameters. 
In the cNMSSM~\cite{Djouadi:2008yj,Djouadi:2008uj} the number of parameters is reduced
by requiring unification of the breaking parameters similar to mSUGRA and
the correct mass of the Z boson. This leaves four parameters. Additionally 
imposing the necessary non--zero minimum value of the Higgs potential favors low m$_0$. Requiring
also the absence of tachyons induces the tri--linear coupling to be negative.
Adding in the LEP constraints, the correct relic density (stau assisted annihilation) 
and the measurement
of $g-2$, the model reduces to lines in the (m$_{1/2}$,A$_0$) plane for low m$_0$
as shown in Figure~\ref{Fig:cNMSSM}. The prediction for the phenomenology at the LHC
is that the stau lepton is the almost stable NLSP. Thus an interesting and a bit unusual 
signature to be studied by ATLAS and CMS.

Recently theoretical studies of the LHC potential for sgluons have been 
performed~\cite{Plehn:2008ae,Choi:2008ub}. 
These $\mathrm{R}=+1$ scalar particles appear when
joining the gauge multiplet with an N=2 type chiral multiplet in order to expand Majorana to Dirac 
gauginos~\cite{Fayet:1975yi}. Examples of such models are the 
$\mathrm{N}=1/\mathrm{N}=2$ hybrid model and the MRSSM~\cite{Kribs:2007ac}. 
The production cross section for sgluon pair production
could be 10~times larger than those of scalar quarks thus being potentially interesting
for rapid discovery/exclusion at the LHC. Experimental signatures are for example 
(cascade via gluinos, then squark--quark followed by a squark to neutralino plus 
quark decay) eight jets with large transverse momenta (plus missing transverse momentum
due to 4~LSPs) or like-sign top quark pairs. The influence of QCD--ISR was studied
and found not to interfere with a sgluon mass measurement once only jets with transverse
momenta of more than 100~GeV are required.

\begin{figure}[htb]
\vspace{-0.75cm}
\includegraphics[width=0.48\columnwidth]{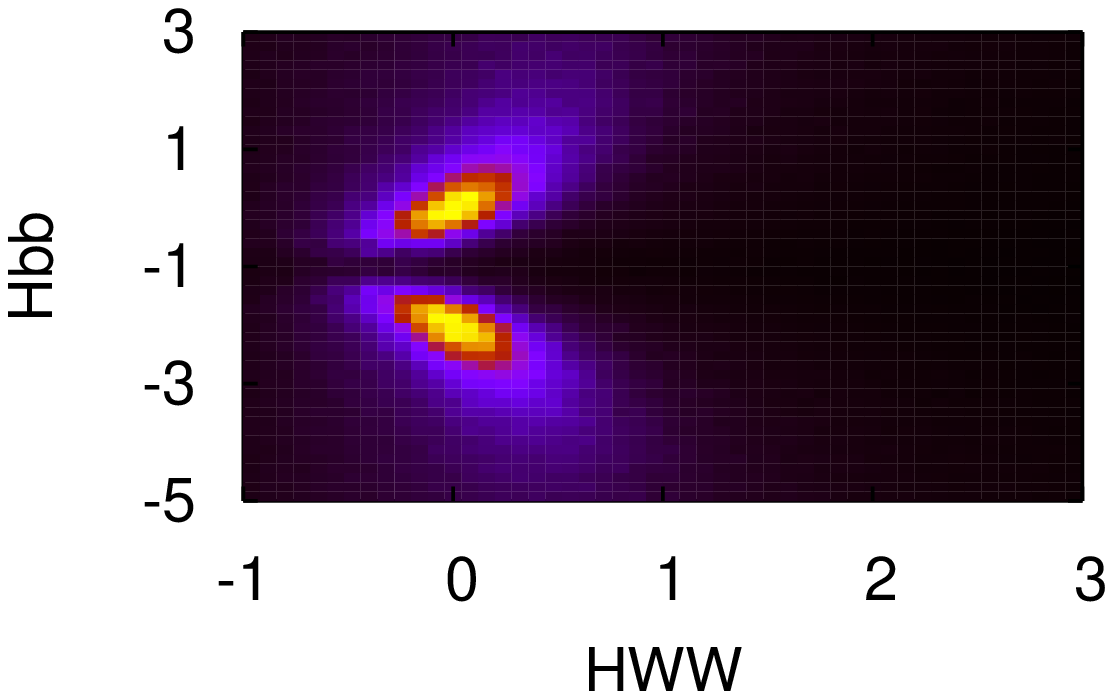}
\includegraphics[width=0.48\columnwidth]{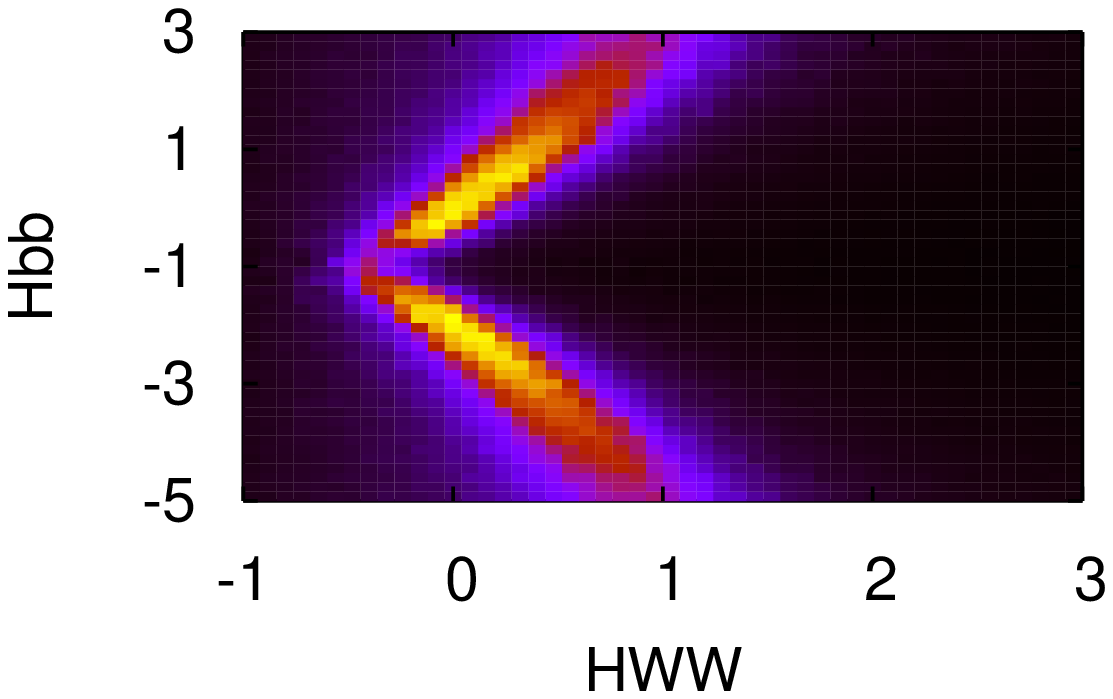}
\vspace{-0.5cm}
\caption{The impact of the use of the subjet analysis is shown for 
the determination of the Higgs boson couplings to b quarks and W bosons.
On the left the full sensitivity is used, on the right, the analysis is excluded
from the determination~\cite{Lafaye:2009vr}.}\label{Fig:Subjet}
\end{figure}

The exact opposite of the exciting scenario of observing large supersymmetric signals
at the LHC would be to observe only a single neutral Higgs boson
with a mass of about 120~GeV. In this case several decay channels would be 
open and similar to the full supersymmetric case, these measurements will be strongly
correlated and theoretical errors will have to be taken into account. 
Thus again the search techniques described in the previous section
can be applied to the determination of the Higgs boson couplings~\cite{Lafaye:2009vr}.
The analysis critically depends on the determination of the Higgs boson coupling to 
b--quarks. In Figure~\ref{Fig:Subjet} the impact of the newly proposed 
subjet analysis~\cite{Butterworth:2008iy} is shown. The impact of this analysis is very important
as the previous golden channel ttH(H$\rightarrow$bb) has turned out to be more difficult
and less sensitive recently~\cite{Ball:2007zza,Aad:2009wy}. Determining the Higgs couplings 
in a scenario close to the SPS1a definition, modifying m$_A$, $\tan\beta$ and A$_t$ 
to move out of the decoupling regime, the log-likelihood is used as estimator
with all correlations:
the Standard Model interpretation the MonteCarloData can be excluded at 90\%~C.L.
for 77\% of the toy experiments, showing the good sensitivity of the Higgs sector
to new physics.

\section{Conclusions}

Supersymmetry remains an 
attractive candidate for new physics at the TeV scale. The LHC could 
provide a wealth of measurements of supersymmetry and is ready 
for difficult scenarios. Collider dark matter property predictions
could be comparable in precision to the relic density measurements 
from WMAP and Planck. The future will tell...

\section*{Acknowledgments}

It is a pleasure to thank the organizers of the session 
\emph{Electroweak physics and beyond the standard model}
for the invitation and the organizers of DIS09 for the wonderful conference. 
Part of the work was developed in the French GDR Terascale (CNRS). 
I am indebted to 
Tilman Plehn, Peter Zerwas, Laurent Serin, Emmanuel Turlay and Claire Adam for their help in the preparation
of the talk and the manuscript.

% ****************************************************************************
% BIBLIOGRAPHY AREA
% ****************************************************************************

\begin{footnotesize}
% IF YOU DO NOT USE BIBTEX, USE THE FOLLOWING SAMPLE SCHEME FOR THE REFERENCES
% ----------------------------------------------------------------------------

% ----------------------------------------------------------------------------

% IF YOU USE BIBTEX,
% - DELETE THE TEXT BETWEEN THE TWO ABOVE DASHED LINES
% - UNCOMMENT THE NEXT TWO LINES AND REPLACE 'Name_Of_Your_BibFile'

%\bibliographystyle{unsrt}
%\bibliography{zerwas_dirk}

\end{footnotesize}

% ****************************************************************************
% END OF BIBLIOGRAPHY AREA
% ****************************************************************************

\end{document}